\RequirePackage[2020-02-02]{latexrelease}
\documentclass{iau}

\usepackage{amsmath}
\usepackage{graphicx}
\usepackage{multirow}

\newcommand{\kms}{km\,s$^{-1}$}
\newcommand{\kmsyr}{km\,s$^{-1}$\,yr$^{-1}$}

\begin{document}

\lefttitle{Brand, Engels, Winnberg}
\righttitle{Patterns: water masers of evolved stars over decades}
\jnlPage{1}{7}

\jnlDoiYr{2023}
\doival{10.1017/S1743921323002144}
%\doival{TBD}

\journaltitle{Cosmic Masers: Proper Motion toward the Next-Generation Large Projects}
\aopheadtitle{Proceedings IAU Symposium}
\volno{380}
\editors{T. Hirota,  H. Imai, K.M. Menten \&  Y. Pihlstr\"om, eds.}

%\title{Variability of water masers in evolved stars on the timescale of decades }
\title{Patterns in water maser emission of evolved stars on the timescale of decades }

\author{Jan Brand$^1$, Dieter Engels$^2$, Anders Winnberg$^3$ }
\affiliation{$^1$ INAF-Istituto di Radioastronomia, Bologna, Italy; email: {\tt brand@ira.inaf.it} \\
%}
%\affiliation{
$^2$ Hamburger Sternwarte, Hamburg, Germany;
%}
%\affiliation{
$^3$ Onsala Rymdobservatorium, Onsala, Sweden}

\begin{abstract}
We present our past and current long-term monitoring program of water masers in the circumstellar envelopes of evolved stars, augmented by occasional interferometric observations. Using as example the Mira-variable U\,Her, we identify three types of variability: periodic (following the optical variation), long-term (years-decades) and short-term irregular (weeks-months). We show there are regions in the maser shell where excitation conditions are favourable, which remain stable for many years. Lifetimes of maser clouds in the wind-acceleration zone are of the order of up to a few years. Much longer lifetimes are found for the peculiar case of a maser cloud outside that zone (as in RT\,Vir), or in some cases where the motion of spectral features can be followed for the entire 2 decade monitoring period (as in red supergiant VX\,Sgr). 
\end{abstract}


\maketitle

\section{Introduction}
In the circumstellar envelopes (CSEs) of O-rich evolved stars, physical conditions are often favourable for the excitation of masers of SiO, H$_2$O and OH, in order of increasing distance from the star. Water masers occur in that part of the CSE where the stellar wind accelerates. As the wind is not homogeneous, masing conditions change with time and with location in the CSE, and the maser emission line profiles are highly variable. Single spectra are not representative; for the study of the evolution of the stellar wind monitoring is therefore essential. \hfill\break\noindent
We have been carrying out monitoring campaigns of water masers in CSEs from 1987 to the present day, with some interruptions. The observations are made with the 32-m antenna in Medicina (1987-2011, 2015 and $>$ 2018) and the Effelsberg 100-m dish (1990-2002). 
Our sample \citep{brand2018} consists of three types of late-type stars: semi-regular variables (SRVs), Mira-variables and red supergiants (RSGs). By making use of occasional interferometric observations (our own, or taken from the literature) we can link peaks in the maser spectral profiles with their spatial counterparts, thus breaking the spatial degeneracy inherent to the line profiles. 

Monitoring data of the SRVs have been discussed in \cite{winnberg2008} and \cite{brand2020}, and papers on the Miras and RSGs are in preparation.  Here we anticipate some results from those analyses.

\section{Variabilities of brightness and timescales}
Figure~1 illustrates the different types of variability we encounter in the monitoring data of Mira variables, using U\,Her as example. The left panel clearly shows a general decline in the integrated flux density of the maser emission over at least two decades. At the same time it is evident that the maser emission responds, with a delay of $\sim$2 months, to the optical variations of the star, as the decreasing flux is modulated with the optical period of 405 days. Superimposed on this are short-time irregular variations. The right panel shows how the emission on the blue side of the stellar velocity gradually becomes less dominant with time, in a somewhat erratic manner; periodic variation is also still recognisable. This change in the blue/red flux ratio indicates differences in excitation conditions in the front- and back caps of the maser shell. This is a consequence of the stellar wind being inhomogeneous, and leads to significant changes in the maser line profile with time. Except for the periodic component, the brightness variability properties of the SRVs \citep{brand2020} are as for the Mira variables.

Our VLA-observations 1990--1992 show that the brightest maser spots are found in the western part of a ring-like distribution in all 4 epochs \citep{winnberg2011}. Maser spot maps (from the literature) taken in the 6.5 years around our observations show the same prevalence. This again points to an inhomogeneous wind, and to  there being regions with favourable excitation conditions that are stable over long times. In this case, the timescale is of the order of a decade.
(Winnberg et al. 2023, in prep.).

\begin{figure}
 \centering
  \includegraphics[scale=.1315]{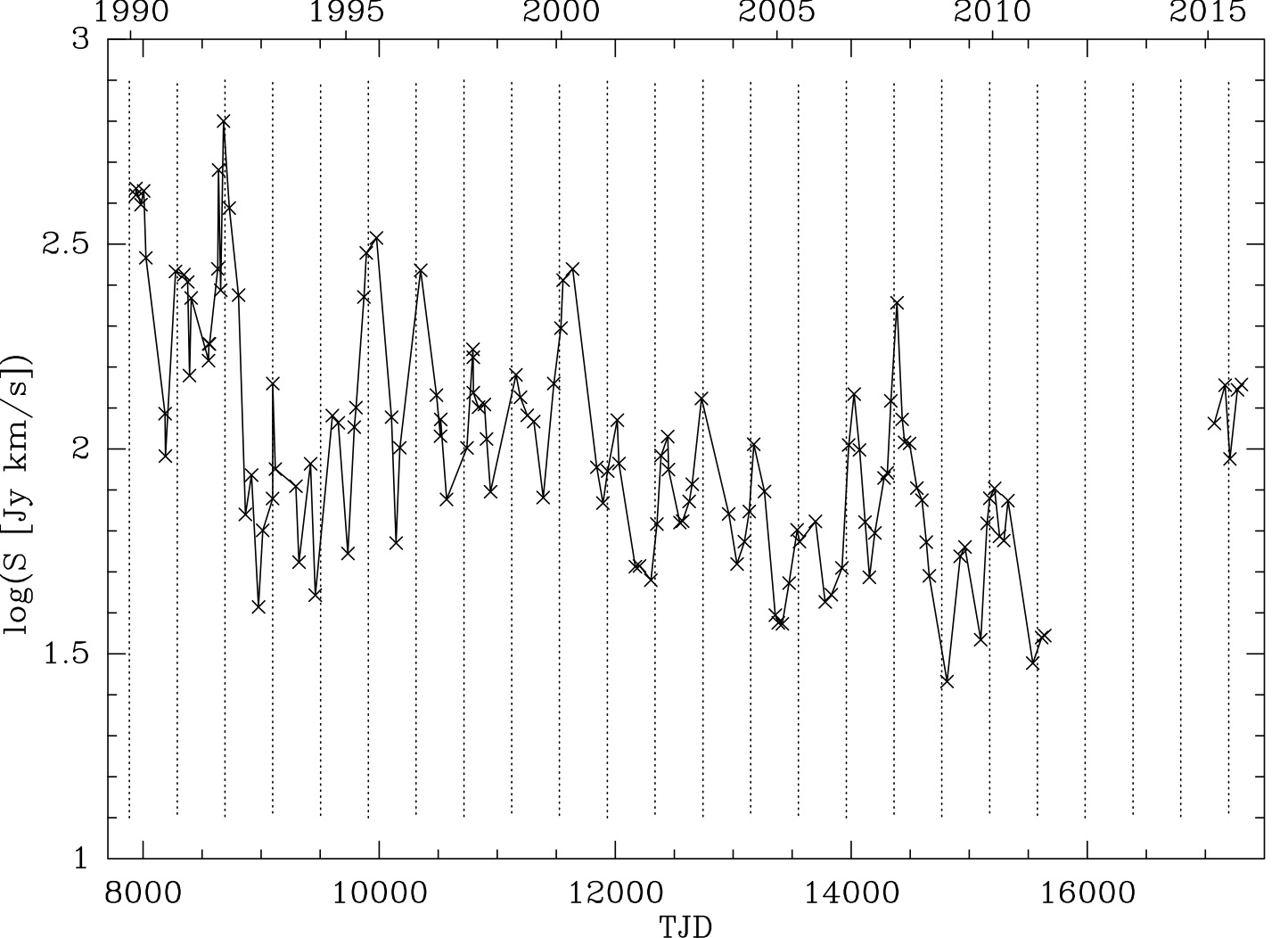}
  \includegraphics[scale=.1315]{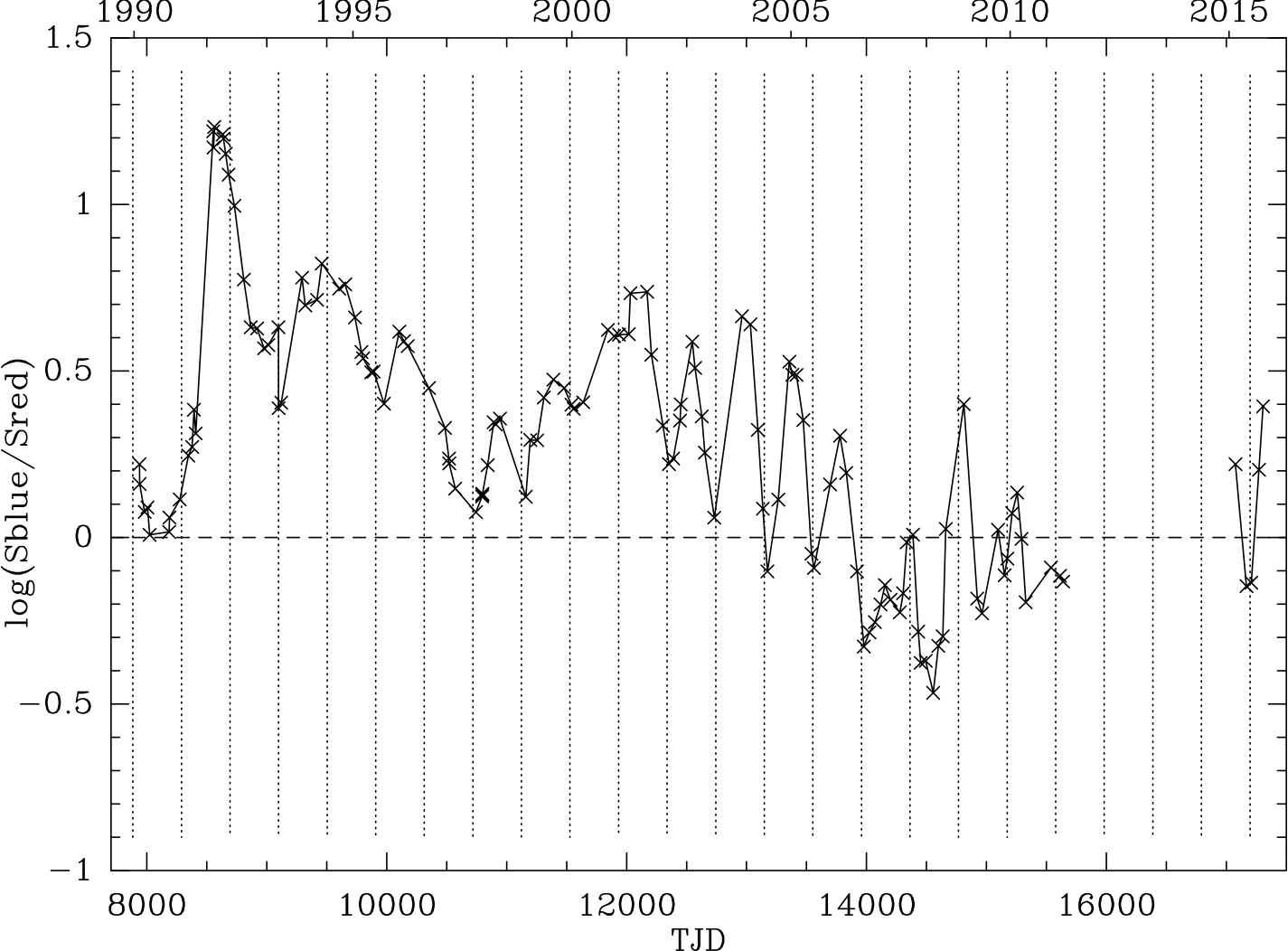}
 \caption{U\,Her. {\it left:}\ Total water maser flux vs. time (TJD=truncated Julian Day=JD-2440000.5). Periodic variability (vertical dotted lines indicate optical maxima), together with overall long-term decline and superimposed short-time irregularities. {\it right:}\ Ratio of total water maser flux on blue and red side (relative to the systemic velocity) of the maser spectra vs. time (from Winnberg et al. 2023, in prep.).
 }
 \label{fig:uher-variability}
\end{figure}

\begin{figure}
    \centering
    \includegraphics[scale=0.13]{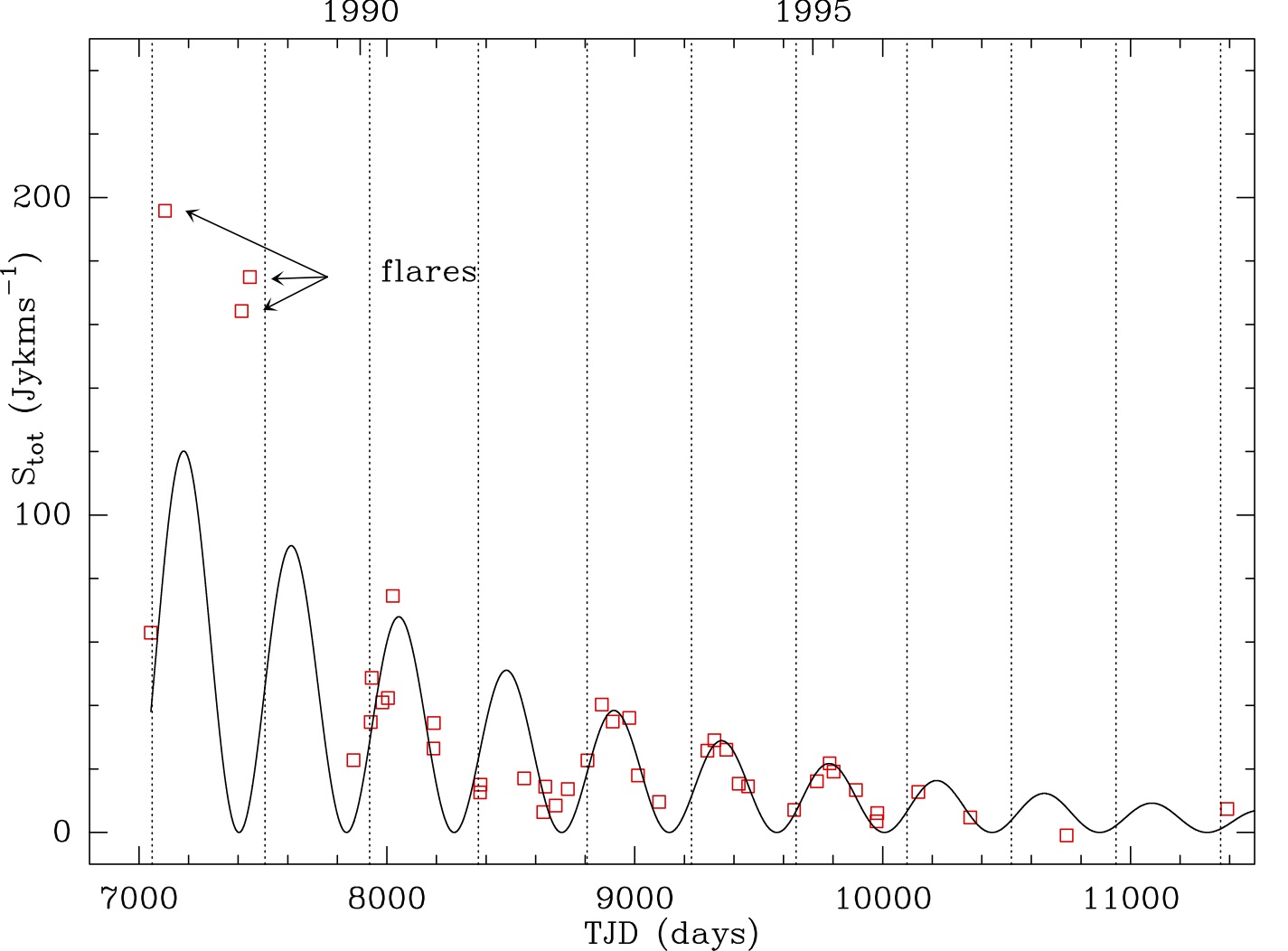}
    \includegraphics[scale=0.13]{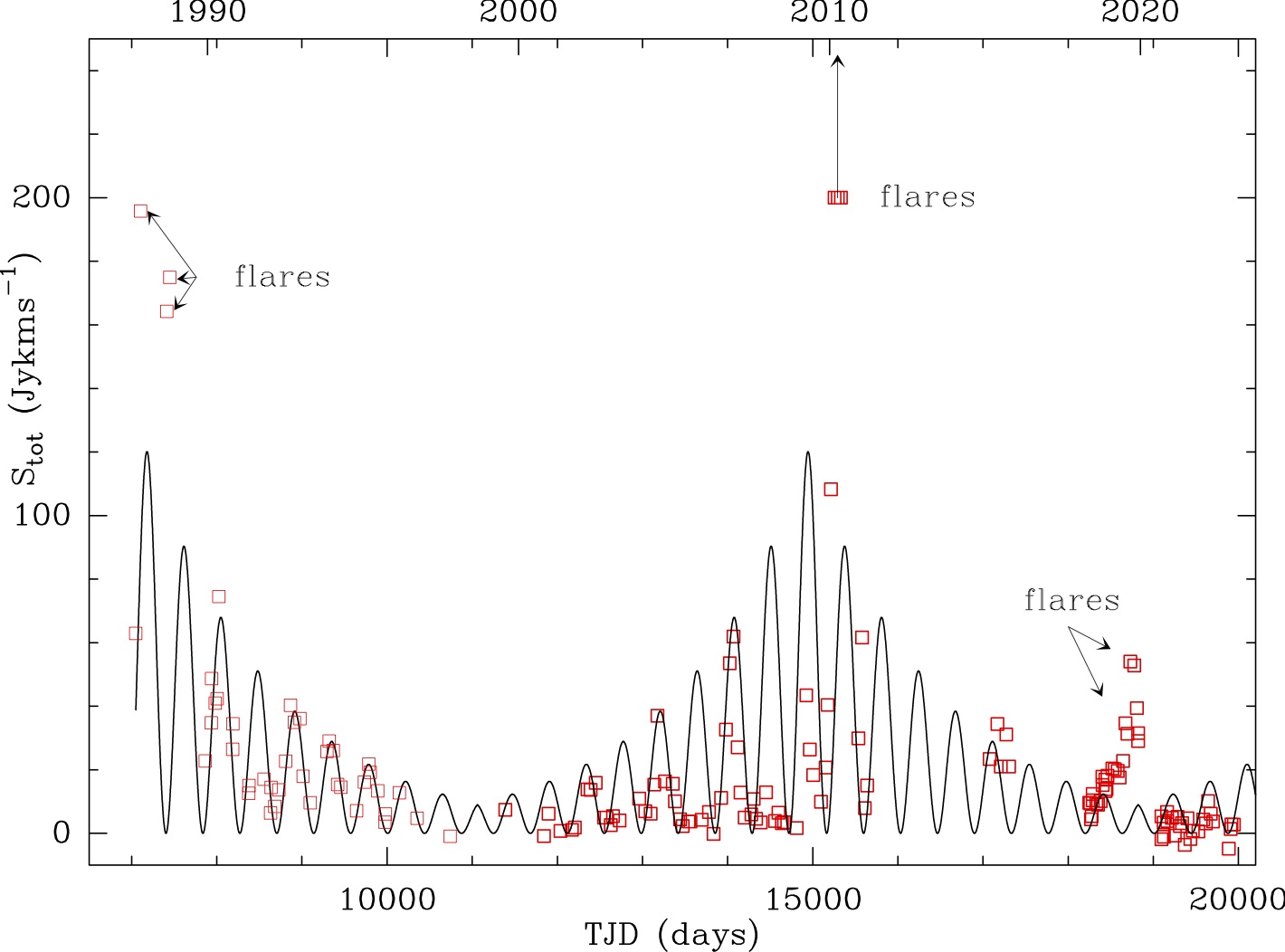}
    \caption{R\,Cas. {\it left:}\ 1987-1999. Dotted lines mark the nominal peaks in optical light. The fitted curve is an exponentially damped sinusoid with a period of 434~days (=P$_{\rm opt}$). {\it right:}\ 1987-2022. }
    \label{fig:rcas-stot}
\end{figure}

\section{R\,Cas: A unique brightness variability pattern}
A unique long-term behaviour of brightness variation is shown by  R\,Cas, which is one of the stars we continue to monitor since 1987. Figure~2 (left) shows the first decade of data (as reported by \citealt{brand2002}), with the fitted damped sinusoid with a period equal to the optical one. Figure~2 (right) shows all data 1987-2022 and this shows a repeating pattern. We interpret this as the result of at least two maser clouds moving through a stationary region in the CSE with favourable conditions for maser excitation (cf. Sect.~2). Emission gradually increases in intensity after entry, reaching a maximum and then slowly decreases as the clouds move out of the region. The entire passage takes about 2 decades; the next maximum is expected in 2029. 

\section{Maser clouds: lifetimes}
A remarkable result of the long-term monitoring program are the almost
constant velocities observed for the major spectral features in SRVs and
Mira variables, as shown for example in the FVt-diagram (Fig.~3, left) of U\,Her. For clarification, the plot on the right shows the velocities of the peaks in the emission profiles, from Gaussian fits. We see how some components are present for some time and then disappear, other lines are visible intermittently, while one or two components are visible more or less the whole time. All show almost no change in velocity. 
Knowing that the water masers originate in an accelerated wind allows one to put an upper limit to the lifetime of a maser spot. 
Assuming a spherically expanding wind, with parameters determined from our maser spot maps, we can calculate the maximum time between two observations before the velocity changes by more than one or two spectral resolution elements ($\lsim$0.6~\kms). For U\,Her this is found to be of the order of 2-4 years. This is thus the maximum lifetime of a maser spot for the velocity to remain constant. On the other hand, in RT\,Vir  we found a maser component with a lifetime of at least 7.5~yr; in this case the maser cloud had likely already moved outside the acceleration zone \citep{brand2020}.

\begin{figure}
    \centering
    \includegraphics[scale=.19]{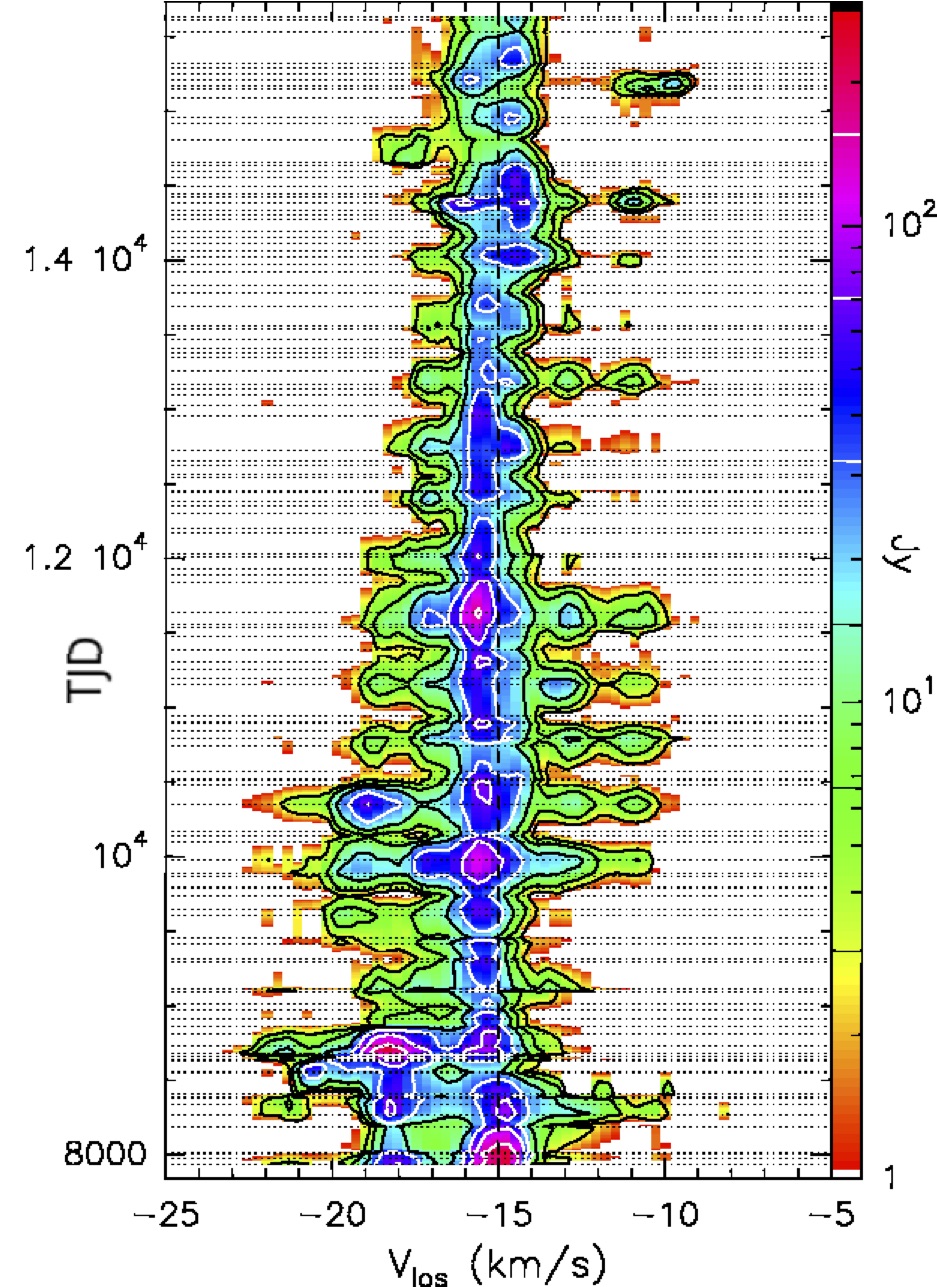}
    \includegraphics[scale=.197]{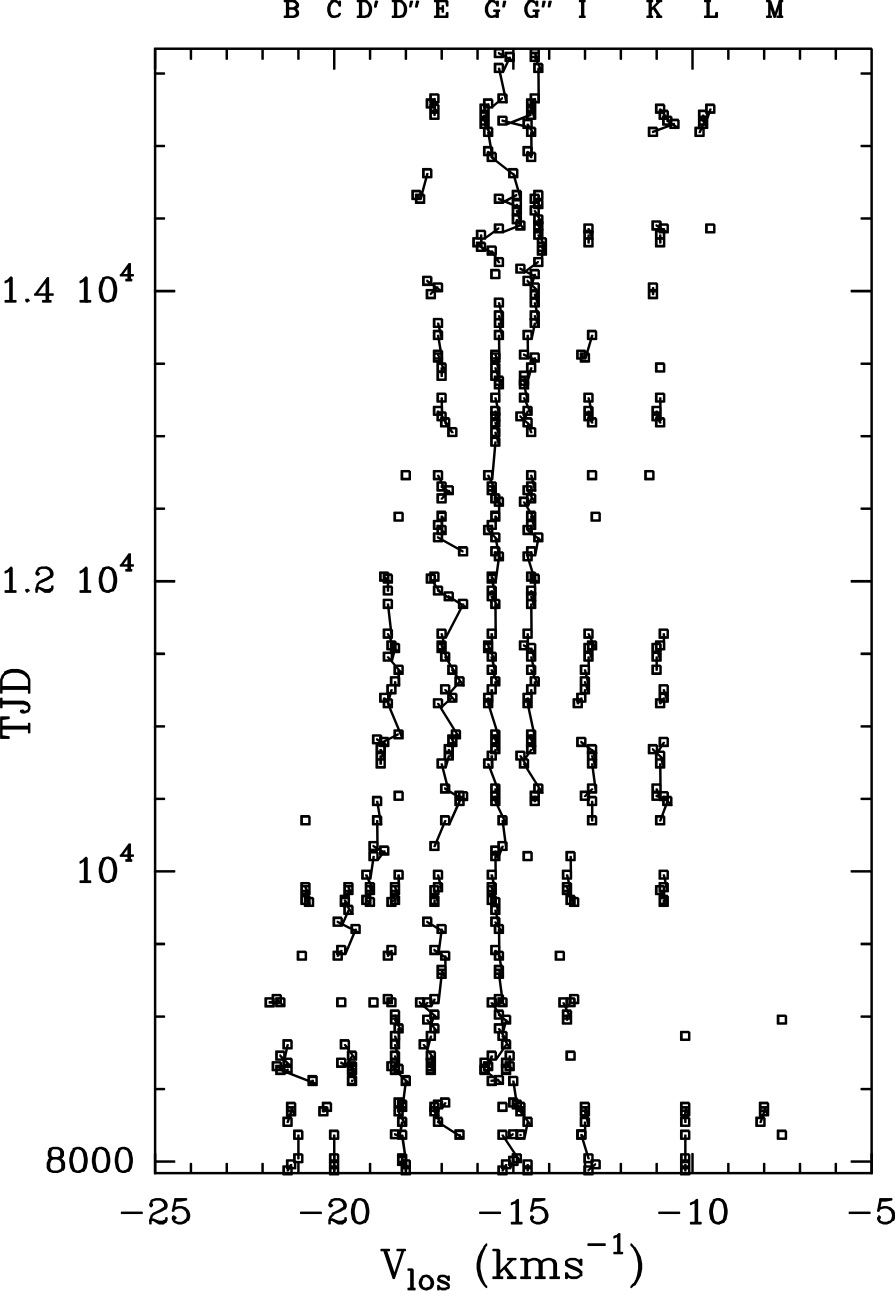}
    \caption{U\,Her 1990-2011. {\it left:}\  Flux density-velocity-time (FVt)-diagram. The stellar velocity is indicated with the vertical dashed line; dotted lines: days with an observation. {\it right:}\ velocities of the peaks of the Gaussian components fitted to the maser spectra, versus time (from Winnberg et al. 2023, in prep.).
    }
    \label{fig:uher-fvt}
\end{figure}

\section{Maser clouds: velocity drifts}
In more extended H$_2$O maser shells with longer lifetimes of the emission regions, velocity drifts are indeed detected. This has been shown for example in the case of IK Tau \citep{brand2018}. Other cases were found among the RSGs. The FVt-diagram of VX\,Sgr (Fig.~4, left) shows that the red edge of the maser emission increases in velocity with time. Figure~4 (right) shows the line-of-sight velocity of the spectral feature defining the red edge (cloud VXSgr-H$_2$O-1987/14), sampled at a few dates. The fitted line indicates an acceleration of $(0.188 \pm 0.008)$~\kmsyr. Using published interferometric maps \citep{murakawa2003} and a spherically expanding CSE-wind model, we find that during our 28~years of monitoring the maser cloud has moved radially outward from 140 to 230~au \citep{engels2021}. 
The case demonstrates that single-dish monitoring with occasional interferometry can indeed trace the movement of maser-emitting clouds in regions with favourable conditions that are blown forward by the stellar wind.

\begin{figure}
    \centering
    \includegraphics[scale=.165]{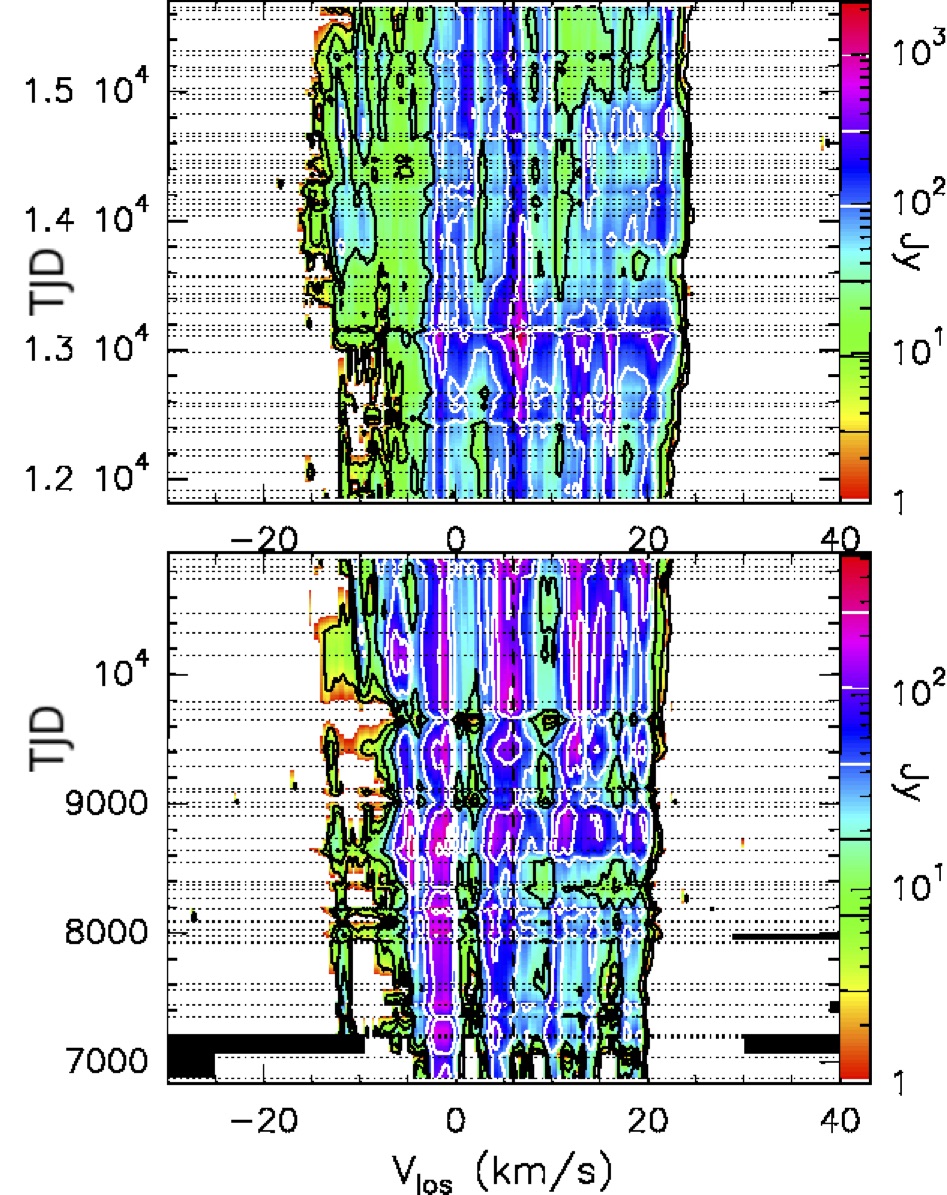}
\includegraphics[scale=.16]{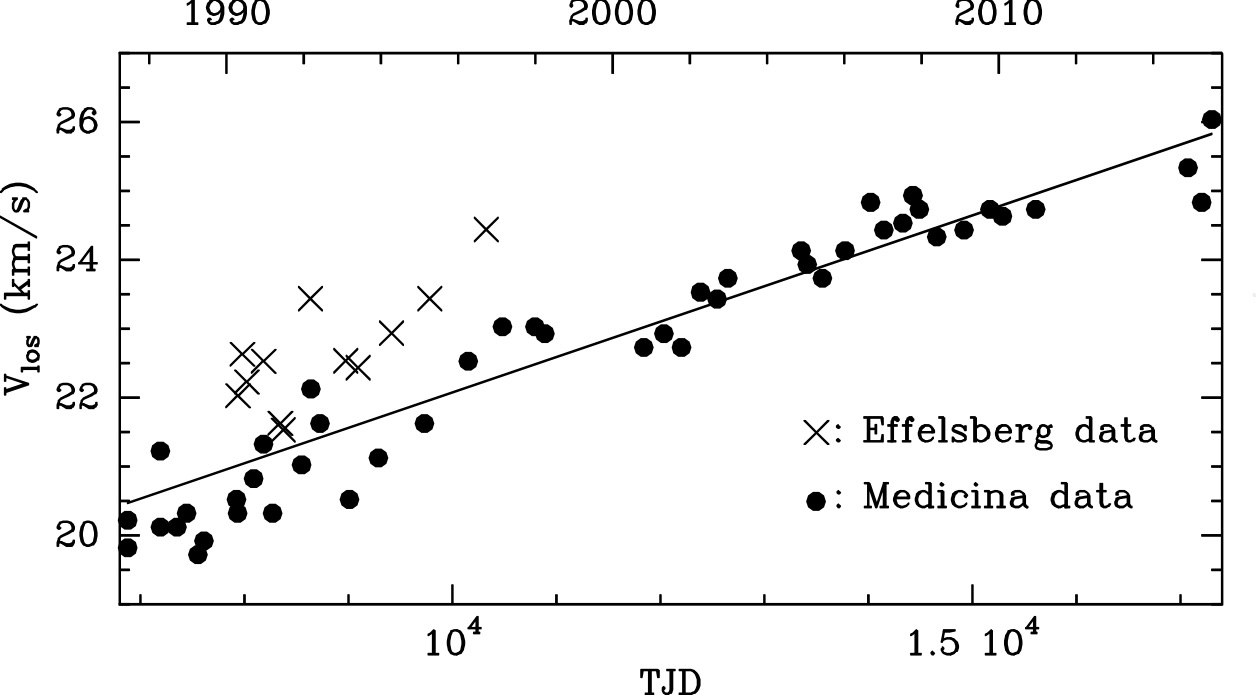}
    \caption{VX\,Sgr. {\it left:}\  FVt-diagram 1987-2011. Note the increase in the velocity of the component defining the red edge of the emission profiles. Dashed and dotted lines as in Fig.~3. Black areas indicate no data. {\it right:}\ Fit to sampled points along the red edge (1987-2015). Drawn line is an rms-weighted fit to the Medicina data, with a slope 0.188 $\pm$ 0.008~\kmsyr\  (from \citealt{engels2021}).}
    \label{fig:vxsgr-fvt}
\end{figure}

\section{Conclusions}
Our monitoring programs show that to trace the structure variations of the stellar wind through water masers, in addition to making occasional interferometric maps, single-dish monitoring with medium-sized radio telescopes is an efficient and practical tool, as these are the only facilities to do time-domain astronomy on timescales of decades. Results from our on-going Medicina Long Project are posted on our dedicated website: https://www.ira.inaf.it/$\sim$brand/Medicina-monitoring.html.

\end{document}